# `SplitStrains`, a tool to identify and separate mixed *Mycobacterium tuberculosis* infections from WGS data

Einar Gabbassov[1,2,*], Miguel Moreno-Molina[3], Iñaki Comas[3], Maxwell Libbrecht[1] and Leonid Chindelevitch[4,*]

## Abstract

The occurrence of multiple strains of a bacterial pathogen such as *M. tuberculosis* or *C. difficile* within a single human host, referred to as a mixed infection, has important implications for both healthcare and public health. However, methods for detecting it, and especially determining the proportion and identities of the underlying strains, from WGS (whole-genome sequencing) data, have been limited. In this paper we introduce `SplitStrains`, a novel method for addressing these challenges. Grounded in a rigorous statistical model, `SplitStrains` not only demonstrates superior performance in proportion estimation to other existing methods on both simulated as well as real *M. tuberculosis* data, but also successfully determines the identity of the underlying strains. We conclude that SplitStrains is a powerful addition to the existing toolkit of analytical methods for data coming from bacterial pathogens and holds the promise of enabling previously inaccessible conclusions to be drawn in the realm of public health microbiology.

## DATA SUMMARY

The authors confirm all supporting data, code and protocols have been provided within the article or through supplementary data files.

Supplementary data files can be found at 10.6084/m9.figshare.14562321.

## INTRODUCTION

Bacterial infections by pathogens such as *Mycobacterium tuberculosis* and *Clostridium difficile* often occur as mixed infections [1, 2], whereby a single patient is infected by several different strains of the same organism. Eukaryotic pathogens such as the main etiological agent of malaria, *Plasmodium falciparum*, can also cause mixed infections [3]. The identification of such mixed infections can be important for reasons including both patient-level decisions [4] as well as public health measures [5]. In the latter setting, if the tracing of the origins of the mixed infection is needed, it may be additionally required to separate the mixed infection into its constituent strains. The separation may also be informative when the mixed infection is hetero-resistant [6], namely, when some, but not all, the strains are resistant to a particular antimicrobial drug. Moreover, a failure to identify the within-host pathogen diversity can lead to misdiagnosing a relapse and reinfection [7]. However, so far, the problem of identifying mixed infections and separating them into their constituent strains has not received a sufficient amount of attention in the literature.

Although older techniques based on the detection of specific regions, such as VNTR (variable-number tandem repeats) [8], are often able to detect such a mixed infection [9], this is not always the case with next-generation sequencing. The main challenge is that the presence of two alternative alleles in a given genomic position may signal a sequencing error as well as the presence of multiple strains. The key distinguishing feature of a mixed infection is the consistency of the fraction of the sample attributable to the sub-dominant strain across most of the variable positions. Thus, depending on the depth of coverage, the similarity between the constituent strains and the proportions in which they are mixed, the problem of detecting and separating mixed strains may vary from straightforward to nearly infeasible.

Several methods for this problem have appeared over the past decade. Eyre *et al.* [2] propose a `Mixed Infection Estimator`, a two-step approach for mixture proportion estimation using a maximum likelihood analysis and mixed strain identification using a custom database. Even though the paper presents results for *C. difficile*, the mixture estimation algorithm can be generalized to other pathogens such as *M. tuberculosis*. This method computes a deviance statistic and uses a threshold value for this statistic to detect mixed infections. As this algorithm was initially designed for *C. difficile* and relies on a custom database of sequences to identify the constituent strains, it could only be used for mixture proportion estimation in our context. More recently, Sobkowiak *et al.* [10] developed `MixInfect`, a method for mixture proportion estimation using a Bayesian model-based clustering technique. This method calculates the ratio of heterozygous calls to total SNPs (single nucleotide polymorphisms) and uses a threshold on this ratio to identify mixed samples. While this algorithm can estimate mixture proportions it does not provide any functionality for resolving the constituent strains. The most recent method, `QuantTB` by Anyansi *et al.* [11], relies on a specially constructed publicly available database of 2166 *M. tuberculosis* assemblies from NCBI [12]. This method provides mixture estimates of WGS samples as well as the identification of strains whose sequence is similar to the ones included in the database. To determine the constituent strains, this method compares the sample to the sequences in the reference database, scoring each of the assemblies. The algorithm then determines how many constituent strains are present in a sample. This approach does not generalize to situations where the underlying strains lack close representatives in the database, which makes its performance highly dependent on the database's representation of the common strains in the relevant local context.

In this paper, we address this problem with a tool called `SplitStrains`, grounded in a rigorous statistical framework. It is based on formulating, for a given set of WGS reads, two alternative hypotheses, namely: the reads belong to a single strain (null hypothesis) or to a mixture of two strains (alternative hypothesis). We then use the EM (Expectation-Maximization) algorithm [13] to estimate the parameters of both hypotheses, and compare their likelihoods to draw a conclusion. As a result, we simultaneously obtain:

- A call to decide whether the sample represents a single (pure) or a mixed infection,
- A likelihood ratio between the alternative and the null hypothesis for the call, and,
- If mixed, the proportion of each constituent strain and a Binary Sequence Alignment Map (BAM) file grouping the reads belonging to each constituent strain.

### Impact Statement

When multiple strains of a pathogenic organism are present in a patient, it may be necessary to not only detect this, but also to identify the individual strains. However, this problem has not yet been solved for bacterial pathogens processed via whole-genome sequencing. In this paper, we propose the `SplitStrains` algorithm for detecting multiple strains in a sample, identifying their proportions, and inferring their sequences, in the case of *Mycobacterium tuberculosis*. We test it on both simulated and real data, with encouraging results. We believe that our work opens new horizons in public health microbiology by allowing a more precise detection, identification and quantification of multiple infecting strains within a sample.

Our results on both simulated and real *M. tuberculosis* data show that `SplitStrains` is effective at identifying mixed infections and continues to perform well even at a relatively low depth of coverage (60×) and low genetic distance (20 SNPs) between strains. Moreover, `SplitStrains` outperforms previously published tools `Mixed infection estimator`, `MixInfect` and `QuantTB` on simulated data. Furthermore, our results show that `SplitStrains` accurately separates the constituent strains provided that their proportions are not too close to each other and they are not too similar. `SplitStrains` is available on GitHub: https://github.com/WGS-TB/SplitStrains.

## METHODS

This part of the paper is organized as follows. First, we briefly describe the datasets used in our analysis. Second, we explain the construction of the feature vector used in our probabilistic model and show how to use it to classify an isolate. Third, we define the Naïve Bayes Classifier for the assignment of reads to strains. Lastly, we show how this approach can be generalized to three or more strains.

We begin by describing the datasets used in our analysis. We report the average number of SNPs relative to the reference genome in the Results section. Here we additionally report the average number of heterogeneous SNPs, defined by a 0/1 in the GT field of the VCF file produced by aligning the sample to the reference genome. We note that the number of heterogeneous SNPs depends on the alignment and variant-calling steps of the pipeline. Therefore, for the *in silico* datasets, this number may be lower than the total number of SNPs added to the reference genome when generating the sample. We report the per-sample statistics in Table S1 (available in the online version of this article).

**Dataset A, *in vitro*.** The 48 mixed *M. tuberculosis* samples presented in [10] are artificially generated *in vitro* by combining the DNA from two clinical cultures of *M.*

*tuberculosis*. The DNA is quantified through spectrophotometry in liquid culture and combined to produce four sets of 12 samples with major strain proportions of 70, 90, 95% (mixed) and 100% (pure). The average number of heterogeneous SNPs is 327.

**Dataset B, *in silico*.** The 60 artificial samples presented in [14] are generated from the standard reference genome for *M. tuberculosis* by substituting randomly chosen alleles at each of 553 genes in an essential core genome MLST scheme (ecgMLST), created by intersecting the set of core genes in an existing scheme with the set of 615 essential *M. tuberculosis* genes [15]. The full dataset contains three pure genomes, with ten samples generated from each one by varying the depth of coverage from 10 to 100 in increments of ten, and three mixed genomes obtained by mixing an additional three pure genomes in pairs, with ten samples generated from each by varying the major strain proportion from 50% to 95% in 5% increments. The average number of heterogeneous SNPs is 2843.

**Dataset C, *in silico*.** For this dataset, generated specifically for this paper, the constituent strains are produced from the H37Rv reference genome. The WGS data is produced by the ART simulator [16] with the following settings:

(1) Profile: HiSeqX PCR free
(2) Read length: 150
(3) Per base sequence quality scores: 20 to 30 on the Sanger/Illumina 1.9 scale.
(4) Quality shift: in order to match the quality of the real data, we shifted the quality scores down by nine to produce relatively uncertain sequences with high sequencing errors.
(5) Depth of coverage: 100 for single-strain and two-strain samples and 150 for three-strain samples.

The dataset consists of eight two-strain mixtures, six three-strain mixtures, and eight single strains. The two-strain mixtures have major strain proportions varying from 50% to 95% in 5% increments, with 55% and 60% omitted. The six three-strain samples have proportions 10:25:65, 15:30:55, 20:35:45, 25:40:35, 30:45:25, and 35:50:15. The average number of heterogeneous SNPs is 136 for the two-strain samples and 583 for the three-strain samples. The simulated reads were aligned back to the reference genome with BWA-MEM [17].

**Dataset E, *in silico*.** For this dataset, generated specifically for this paper, the constituent strains are produced from the same H37Rv reference genome with $N \in \{10, 15, 20, 25\}$ random base substitutions. This yields four subsets that contain single and two-strain samples. The first subset contains eight single strain and eight two-strain samples with proportions varying from 50% to 95% in 5% increments, with 55% and 60% omitted. All the samples in the first subset have ten SNPs relative to H37Rv, and since these SNPs are chosen independently at random, the two-strain samples are 20 SNPs apart. The remaining sample subsets have the same proportions, but more SNPs. The genetic distances between mixed strains in each of the subsets are thus 20, 30, 40 and 50 SNPs, respectively. The WGS data is produced by the ART simulator with the same settings used to generate Dataset C except for the depth of coverage, which is set to 60 for all the samples.

## Required input data

The `SplitStrains` pipeline uses the BWA-MEM tool, which makes use of paired-end information to produce the alignment. The current pipeline only keeps those pairs for which both elements have been mapped, in order to reduce the possibility of errors. Hence, `SplitStrains` only requires a BAM file of the paired-end data, a reference genome, and optionally, a generic feature format (GFF) file. We say that a sample represented in the BAM file contains a **single strain** if it is pure, a pair of strains called **major strain** and **minor strain** if it is a mixture of two strains, or **multiple strains** in the case of a mixture of more than two strains.

## Data pre-proccessing

All datasets are pre-processed with Trimmomatic [18]. The Trimmomatic settings are:

ILLUMINACLIP=TruSeq3-PE-2,

SLIDINGWINDOW=4 with trimming threshold=16,

Leading=10,

Trailing=10,

Minlen=40.

## Feature vector construction

The BAM file contains the alignment information for each individual read of a sequenced organism. We first convert the BAM file into a pileup format using the `pysam` library [19]. This pileup format summarises the alignment information for each individual base of the reference genome. We then construct a per-base feature vector defined as follows:

$$\boldsymbol{x_i} := (p_A^{(i)}, p_C^{(i)}, p_G^{(i)}, p_T^{(i)}; d^{(i)}), \qquad (1)$$

where $p_b^{(i)} \in [0, 100]$ for $b \in \{A, C, G, T\}$ is the percentage of base $b$ at position $i$, so that $\sum_{b \in \{A, C, G, T\}} p_b^{(i)} = 100$, and $d^{(i)}$ is the total depth (number of aligned reads) at this position. Note that if $p_b^{(i)} > 0$ for more than one $b$, there could be a SNP at position $i$. In practice, at most two of the $p_b^{(i)}$'s are non-zero most of the time. In the absence of sequencing errors, we expect exactly one of the $p_b^{(i)}$'s to equal 100 for every in the case of a single strain. On the other hand, in the case of a mixed sample that has a major strain at a proportion $p \geq \frac{1}{2}$ and a minor strain at a proportion $1 - p$, we expect to see $p_{b_i}^{(i)} = p$ and $p_{b_i'}^{(i)} = 1 - p$, where $b_i \neq b_i'$ for sufficiently many positions $i$.

Fig. 1 shows an example with two adjacent positions, $i$ and $j$. There are $n = 8$ reads supporting position $i$, six of them containing an $A$ and the remaining two containing a $T$. The depth is $d^{(i)} = 8$ and the percentages of bases A and T at are $p_A^{(i)} = 75$ and $p_T^{(i)} = 25$, respectively. The adjacent position

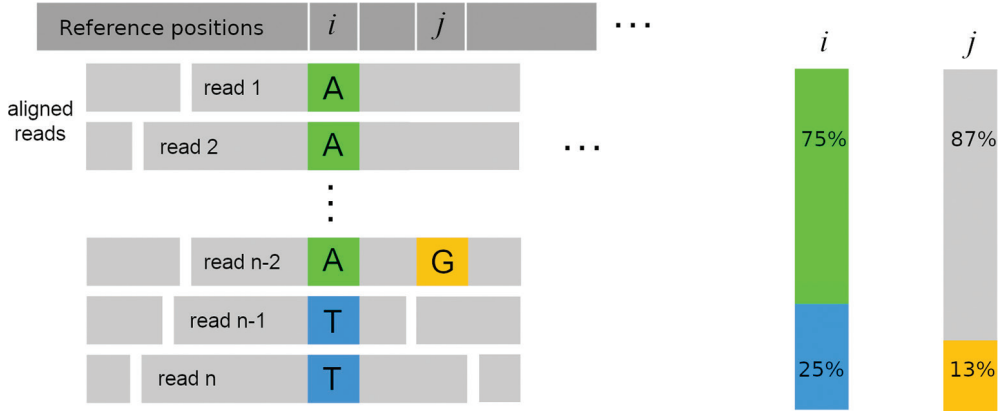

**Fig. 1.** An example read alignment and the corresponding feature vector.

*j* has the same depth and a *G* in one of the reads, with the remaining reads containing a *C*. They are summarized by the feature vectors $\boldsymbol{x}_i = (75, 0, 0, 25; 8)$ and $\boldsymbol{x}_j = (0, 87.5, 12.5, 0; 8)$, respectively.

We empirically observed that sites with a relatively low depth of coverage contained reads that could not be reliably aligned by BWA-MEM (i.e. they had poor alignment quality scores). For this reason, we chose to filter out any site $\boldsymbol{x}_i$ with depth coverage $d^{(i)}$ below $k = 70\%$ of the mean depth of coverage. Additionally, we use a GFF file based on the *M. tuberculosis* reference genome (NC 000962.3), but with the mobile and PE/PPE genes removed. This user-customizable GFF file ensures that `SplitStrains` analyses only annotated gene regions excluding mobile and PE/PPE genes, as the latter are known to be highly repetitive and produce unreliable alignment and variant calling results [20].

### Detecting mixed samples

We test two hypotheses: the null hypothesis ($H_0$), which states that there is a single strain in the data, and an alternative hypothesis ($H_1$), which states that there are two strains with proportions $p$ and $1 - p$, respectively. Our data $D$ consists of all the feature vectors $\boldsymbol{x}_i$ described above.

The likelihood of the data $D$ under $H_0$ is based on the fact that, for every position $i$, the feature vector $\boldsymbol{x}_i$ can only have a non-zero percentage at one base $b$, that is $p_b^{(i)} = 100$. That is, under the null hypothesis, the maximum likelihood estimator of the underlying base given a single strain would always be the most frequently observed nucleotide. However, we also allow sequencing errors to occur with probability $\epsilon_0$. Therefore, using the notation $d_i$ for the number of reads that map to position $i$, and $k_i$ for the number of those reads that end up with the most frequent base, we have

$$P(D \mid H_0) = \prod_i \binom{d_i}{k_i} \epsilon_0^{d_i - k_i}(1 - 3\epsilon_0)^{k_i}, \text{ with } k_i := d_i \cdot \max_b p_b^{(i)}. \quad (2)$$

where the last term in the product arises from the fact that a sequencing error can occur in three different ways, so the probability of getting the correct base is $1 - 3\epsilon_0$.

Under $H_1$ we assume that there are two strains with proportions $p$ and $1 - p$, and sequencing errors which occur with probability $\epsilon_1$. We now use $n_M^{(i)} := d_i \cdot \max_b p_b^{(i)}$ and $n_m^{(i)} := d_i \cdot \max_{b' \neq b} p_{b'}^{(i)}$ to denote the counts of the most and second most frequent bases at position $i$, and $n_e^{(i)} := d_i - (n_M^{(i)} + n_m^{(i)})$ to denote the count of the remaining bases.

Repeating the above analysis gives us the following expression:

$$P(D \mid H_1) = \prod_i \binom{d^{(i)}}{n_M^{(i)}, \quad n_m^{(i)}, \quad n_e^{(i)}} \left(p(1 - 3\epsilon_1) + (1 - p)\epsilon_1\right)^{n_M^{(i)}}\left((1 - p)\right.$$
$$\left.(1 - 3\epsilon_1) + p\epsilon_1\right)^{n_m^{(i)}}\epsilon_1^{n_e^{(i)}}, \quad (3)$$

where

$$\binom{d^{(i)}}{n_M^{(i)}, \quad n_m^{(i)}, \quad n_e^{(i)}} := \binom{d^{(i)}}{n_M^{(i)}}\binom{d^{(i)} - n_M^{(i)}}{n_m^{(i)}} \quad (4)$$

is the trinomial coefficient, which generalizes the binomial coefficient.

Using the likelihood ratio test, we can now formulate the following condition:

if $-2\left(\log(P(D \mid H_0)) - \log(P(D \mid H_1))\right) < c$, then select $H_0$, otherwise $H_1$. (5)

The threshold value $c$ is defined based on the significance level $\alpha$ using the $\chi^2$ distribution with one degree of freedom. Finally, in order to evaluate Equation (5) we estimate the parameters $\epsilon_0$ for $H_0$ and $p, \epsilon_1$ for $H_1$ using the well-established Truncated Newton constrained optimization algorithm [21]. In both cases we estimate a set of parameters that maximize the likelihood of the data conditional on the hypothesis, i.e.

we perform maximum likelihood estimation on the corresponding parameters.

## Read assignment with two strains

In order to obtain a read assignment $r$ to a strain, we wish to compute the probability of a read $r$ belonging to the major strain $M$ and to the minor strain $m$. Let $\Pr[r \in M]$ and $\Pr[r \in M]$ denote the respective probabilities. A read often supports multiple SNPs at the same time, say, at positions $i_1, i_2, \ldots, i_n$; we let $C_r = \{i_1, i_2, \ldots, i_n\}$ be the set of such positions for a read $r$.

For each $i \in C_r$ we get the counts of each base $b \in \{A, C, G, T\}$ from the feature vector $x_i$. Let $x^{(i)}$ denote the count of the base $r_i$ found in the read $r$ at position $i \in C_r$. We then define the following condition:

$$\text{if } \frac{\Pr[r \in M | x^{(i)}, i \in C_r]}{\Pr[r \in m | x^{(i)}, i \in C_r]} \geq 1, \text{ then } r \in M, \text{ otherwise } r \in m. \quad (6)$$

We now analyse Equation (6) under two alternative models - a binomial one and a Gaussian one with equal variances (which would be the case if the Gaussian model was approximating the binomial one because $\sigma_1^2 = d^{(i)}p(1-p) = d^{(i)}(1-p)p = \sigma_2^2$. We show that both of these models simplify to a majority vote of the variants present inside a read, where the $i$—th variant's number of votes for a strain equals to its frequency in the feature vector, $x_i$. However, these votes are unweighted in the binomial model, while in the Gaussian model, each variant's total number of votes is normalized to one by dividing it by the depth of coverage $d^{(i)}$.

Suppose that the proportions of the major and the minor strains are $p$ and $1 - p$, respectively, as inferred from the mixture model. Since $p$ is the proportion of the major strain, we may assume that $p \geq \frac{1}{2}$. To further simplify the computation, we assume that each $x_i$ is independent. Then the left-hand side of Equation (6) can be expressed as a ratio of products:

$$\text{if } \frac{\prod_{i \in C_r} \binom{d^{(i)}}{x^{(i)}} p^{x^{(i)}} (1-p)^{d^{(i)} - x^{(i)}}}{\prod_{i \in C_r} \binom{d^{(i)}}{x^{(i)}} (1-p)^{x^{(i)}} p^{d^{(i)} - x^{(i)}}} \geq 1, \text{ then } r \in M, \text{ otherwise } r \in m. \quad (7)$$

Simplifying the left-hand side further, we arrive at a simple condition, independent of $p$:

$$\text{if } \sum_{i \in C_r} (2x^{(i)} - d^{(i)}) \geq 0, \text{ then } r \in M, \text{ otherwise } r \in m. \quad (8)$$

Therefore, by applying Equation (8) we can classify whether a read $r$ belongs to the major or the minor strain. In the case of a perfect tie, we assign the read to the major strain.

If, instead of the derivation above, we use the Gaussian probability density functions $f(x | \mu, \sigma_1)$ and $f(x | 1 - \mu, \sigma_2)$ to model the ratios in Equation (6), and further assuming equal variances $\sigma = \sigma_1 = \sigma_2$, we arrive at a condition very similar to that in Equation (8):

$$\frac{\prod_{i \in C_r} f(x^{(i)}/d^{(i)} | \mu, \sigma)}{\prod_{i \in C_r} f(x^{(i)}/d^{(i)} | 1-\mu, \sigma)} \geq 1 \text{ iff } \sum_{i \in C_r} (2x^{(i)}/d^{(i)} - 1) \geq 0. \quad (9)$$

Although similar, Equations (8) and (9) can lead to opposite conclusions when the depth of coverage $d^{(i)}$ varies between the positions occurring in a read; for instance, when a read $r$ contains two biallelic positions, with the first one occurring in six reads of which five (including $r$) agree with the major strain, and with the second one occurring in nine reads of which two (including $r$) agree with it, then the binomial model would assign this read to the minor strain (since $2 (5 + 2) = 14 < 15 = 6 + 9$), while the Gaussian model would assign it to the major strain (since $5/6 + 2/9 = 19/18 > 1$). On the other hand, it is easy to see that if the coverage is uniform, i.e. $d^{(i)} = D$ for every position in the read, then Equation (8) and Equation (9) are equivalent. Hence, it is possible to use either the binomial or the Gaussian distribution in Equation (6), with identical assignments when we assume uniform coverage and equal variances and normalize the means to add up to one. In our implementation we choose not to constrain the variances to be equal, and apply the left-hand side of Equation (9) with the empirically fitted mean and variance, but with the fitted means normalized by their sum prior to the computation. This leads us to the more general situation, where we have $n > 2$ strains in a sample.

## Binomial and Gaussian mixture models for multiple strains

In this subsection we describe a probabilistic model for proportion estimation and read assignment in the case of multiple strains.

In order to build the Mixture Model, we use the information in the feature vectors $x_i$ to construct a matrix $X$ as follows

$$X := [p_A^{(i_1)}, p_C^{(i_1)}, p_G^{(i_1)}, p_T^{(i_1)}, \ldots, p_A^{(i_N)}, p_C^{(i_N)}, p_G^{(i_N)}, p_T^{(i_N)}], \quad (10)$$

where $\{i_n\}_{n=1}^N$ is a strictly increasing sequence of integer indices (i.e. $i_1 < i_2 < \ldots < i_N$) of variable positions. Now, let $K$ be the number of strains in a sample. We denote the unknown proportions of each strain as $\mu_1, \ldots, \mu_K$, the standard deviation of each proportion as $\sigma_1, \ldots, \sigma_K$. and the weight of each mixture model component, as $w_1, \ldots, w_K$. Define the parameter vector $\boldsymbol{\theta} := \{\mu_1, \sigma_1, \ldots, \mu_K, \sigma_K, w_K\}$; then the mixture model has the following form:

$$p(\boldsymbol{\theta} \mid X) = \sum_{k=1}^K w_k f(x \mid \mu_k, \sigma_k), \quad (11)$$

where $f(x | \mu_k, \sigma_k)$ is the density function of a binomial or Gaussian distribution.

For a given $X$, we use the well-established EM (expectation maximization) algorithm to learn $\boldsymbol{\theta}$.

Once the model has been learned (Fig. 2), it is possible to proceed to the assignment of each read to a strain via Naïve Bayes classification:

$$\mathbf{p}_k = \Pr[r \in S_k \mid p_{r_i}^{(i)} \text{ for } i \in C_r] \propto w_k \prod_{i \in C_r} f(p_{r_i}^{(i)} \mid \mu_k, \sigma_k). \quad (12)$$

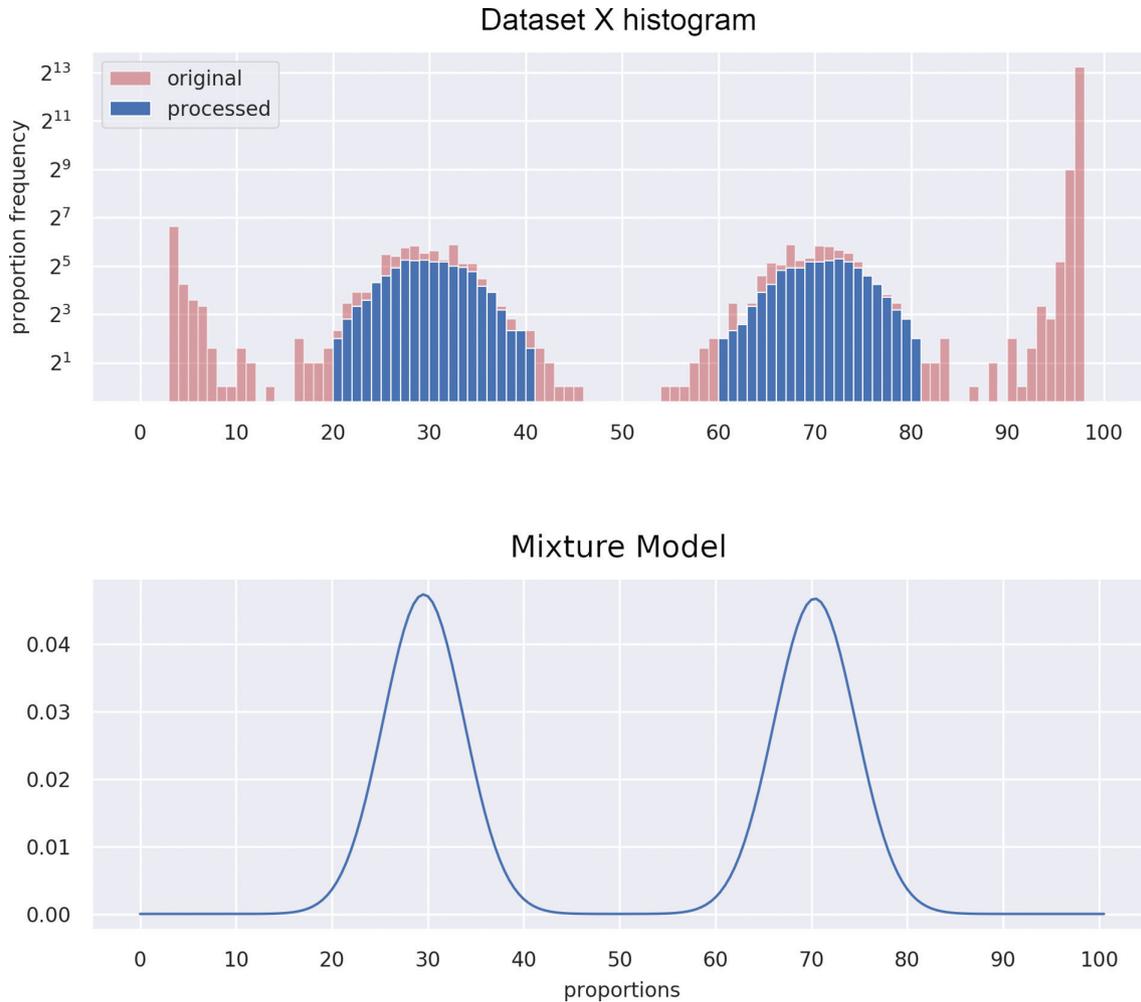

**Fig. 2.** Example of a simulated mixed sample, the data $X$, containing a major and a minor strain with respective percentages $\mu_1 = 70, \mu_2 = 30$. The red bars represent filtered out sites $\boldsymbol{x_i}$ with depth $\boldsymbol{d^{(i)}}$ less than $\kappa = 70\%$ of the average depth of coverage.

Finally, we use a maximum a posteriori assignment: $r \in S_j$ if and only if $\boldsymbol{p_j} = \max \{\boldsymbol{p_k}\}$.

**Computational settings**

All our computations are performed on a 64-bit Ubuntu Linux computer with eight CPU cores and 16 GB of RAM. The entire pipeline's running time ranges from 2 minutes to 30 minutes per sample, depending on the settings and the depth of coverage.

## RESULTS

For simplicity, a WGS sample will be called *pure* if it contains a single strain of the sequenced organism and *mixed* otherwise. The `SplitStrains` algorithm classifies a sample as being pure or mixed. If the sample is classified as mixed, the algorithm detects the proportion of each strain and separates the reads according to which strain they belong to. In order to accomplish this, the algorithm proceeds through three stages.

First, `SplitStrains` uses the sample's SNPs to infer the parameters of a Gaussian or binomial Mixture Model (GMM), which identifies the number and the proportions of the constituent simple strains. The likelihood ratio statistic produced in the process provides a rigorous quantification of the confidence about its status as a pure or mixed sample. The algorithm then uses the model's estimated parameters in a Naïve Bayes classifier to assign each read that contains variable positions to one strain. Finally, it produces a BAM file for each constituent strain. Note that any read that does contain any variable positions does not get classified, and instead gets included in all the BAM files produced. The process is shown in Fig. 3.

In principle, it would be possible for `SplitStrains` to partition the SNPs instead of the reads. However, we choose to partition the reads for two reasons. First, there is generally more information to classify a read than there is to classify a SNP, because a single read can contain multiple SNPs, as we explain in the Methods. Second, an assembly program (either reference-based or *de novo*) can be applied directly to the BAM files representing each strain, which can make the downstream analysis more accurate in cases where the variation present in some of the strains is not limited to SNPs, but also includes indels or copy number variants.

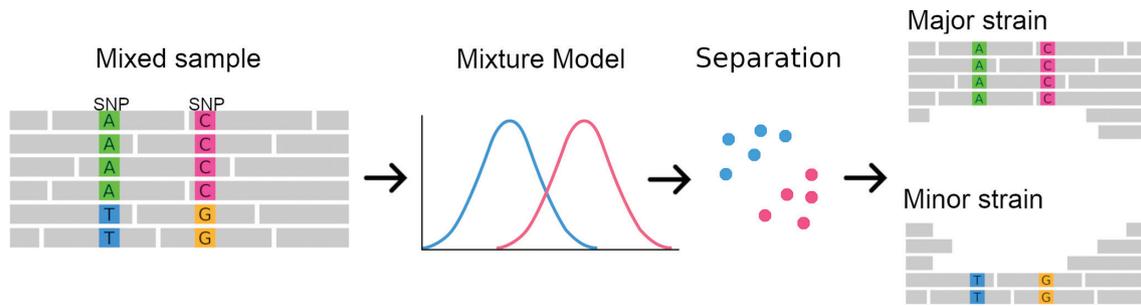

**Fig. 3.** `SplitStrains` workflow overview.

Although `SplitStrains` is primarily designed for samples containing one or two strains, we also apply it to situations, rarely seen with *M. tuberculosis* but more common with other bacteria, where three strains are present in a single sample. We note that our likelihood ratio statistic is calibrated to distinguish between the pure and the mixed case but is not in general sensitive enough to distinguish between two-strain and three-strain samples. For this reason, we consider our mixed sample detection to be correct if it classifies a mixed sample as mixed, regardless of the number of underlying strains.

To evaluate the strain proportion estimation in a uniform way, we consider the accuracy of the proportion estimation for the major strain (which, in the case of a two-strain sample, is equivalent to the accuracy for the minor strain). Lastly, to evaluate the assignment of reads to constituent strains, we separately analyse the samples with two and three strains, and explicitly specify the number of constituent strains as two or three.

We evaluate the performance of the `SplitStrains` algorithm by quantifying the accuracy of its mixture proportion estimation and strain separation on several datasets:

- **Dataset A** 48 *in vitro* samples, including 12 pure and 36 artificially mixed two-strain samples with known strain proportions, published together with a previous method, `MixInfect` [10]. The samples have an average of 1249 SNPs relative to the reference genome. This dataset is the most realistic representation of mixed infections since it is generated *in vitro* by combining the DNA from two *M. tuberculosis* cultures.
- **Dataset B** 60 *in silico* samples, including 30 pure and 30 mixed two-strain samples with known strain proportions,

published together with a study on whole-genome MLST schemes from our group [14]. The samples have an average of 6375 SNPs relative to the reference genome. This dataset is the second most realistic dataset since it is based on previously observed alleles of genes and includes a wide range of different mixture proportions.

- **Dataset C** 22 *in silico* samples, of which eight are pure and 14 are mixed (eight with two and six with three constituent strains) produced specifically for this work. The mixed samples are derived directly from the reference genome by independently adding 100 SNPs per strain in the two-strain samples, and 300 SNPs per strain in a three-strain sample. This is the least realistic dataset, and it is designed to test `SplitStrains`'s ability to generalize to data with one, two or three strains and a wide range of mixture proportions.
- **Dataset D** 59 *real* samples chosen among those collected in population-level surveys in Azerbaijan, Bangladesh, Belarus, Pakistan, Philippines, South Africa (Gauteng and Kwazulu Natal provinces) and Ukraine, collected between 2009 and 2014 [22], with additional samples from a large-scale whole genome sequencing study conducted in Malawi [23]. The samples have an average of 1248 SNPs relative to the reference genome. This dataset contains real samples, and their true label (pure or mixed) is unknown.
- **Dataset E** 64 *in silico* samples, 32 pure ones and 32 mixed ones with two strains each, used to test the method at a low depth of coverage (60) and a low genetic distance. The 32 mixed samples have one of eight known major strain proportions, and for each proportion, the samples are derived from the reference genome by independently adding 10, 15, 20, or 25 SNPs per strain. This is a calibration dataset designed to test `SplitStrains`'s ability to detect mixed infections with very short genetic distances.

The analysis starts by classifying a given sample as mixed or pure. It uses the likelihood ratio test (LR) to compare the single strain (null) and the multiple strains (alternative) hypotheses. We use the LR statistic to guide the decision process. The algorithm also reports the estimated mixture proportions. If the sample is called mixed, the algorithm further classifies each read containing one or more variants as belonging to a specific strain. Note that this assignment can be extended to a full strain assembly as is frequently done in metagenomics [24], but we do not explicitly do so to focus on our contribution and avoid the complications due to the choice of a suitable assembly algorithm [25]. We now discuss the performance of `SplitStrains` on each of the datasets, omitting the real Dataset D which cannot be used for evaluation due to the absence of "gold standard" information.

### Mixed sample detection

`SplitStrains` is consistently able to detect mixed infections across all the datasets, which suggests its robustness to different numbers of SNPs, depths of coverage, minor strain proportions, and synthesis methods.

**Dataset A** `SplitStrains` correctly classifies all 12 pure samples when the significance level threshold set to α = 0.05. Then 23/24 mixed samples with major strain proportions of 70 and 90% are also correctly classified. 11/12 samples with a major strain proportion of 95% are misclassified as pure; however, by increasing the significance level threshold to α = 0.1, `SplitStrains` correctly classifies 8/12 of these samples, while the classification results for the 24 mixed samples remain unchanged. In total, 43/48 samples (90%) get correctly classified with α = 0.1.

**Dataset B** `SplitStrains` classifies 29/30 mixed samples as mixed, the exception being a sample with a major strain proportion of 95%. However, using α = 0.1 instead of α = 0.05 allows for the correct classification of all 30 mixed samples. Then 27/30 pure samples get classified as pure with α = 0.1, and the three misclassified samples have a very low average depth of coverage (10). In total, 57/60 samples (95%) get correctly classified with α = 0.1.

**Dataset C** `SplitStrains` correctly detects 13/14 mixed samples using α = 0.05, including 6/6 samples with three constituent strains. The misclassified sample is the two-strain sample with a major proportion of 95%. All 8 pure samples are correctly classified as pure. In total, 21/22 samples (95%) get correctly classified as pure with α = 0.05.

**Dataset D** We also applied `SplitStrains` to a real dataset containing 59 samples. A preliminary analysis used a reference mapping pipeline with variant calling optimized for curated databases [26], and declared 27 of the samples as mixed. `SplitStrains` classifies 24/27 of these samples as mixed with α = 0.05, and classifies the remaining 35 samples as pure, demonstrating a concordance of 56/59 (95%) with the reference mapping pipeline that makes extensive use of database information.

### Mixture proportion estimation

In those datasets where the true proportion of the major strain is known, we can compare that proportion to the one inferred by `SplitStrains`, conditional on its correctly classifying the sample as mixed. The correctly classified mixed samples have a maximum deviation from the true solution of 11%; the worst case occurs for a sample with true major proportion of 95%, which is estimated as 84% by `SplitStrains`. In general, the estimation is accurate up to a 90% major strain proportion but starts to decrease as this proportion approaches 95% (Fig. 4). We measure the deviation of all the estimates from their true values using the Root Mean Square Error (RMSE). Averaged across all three datasets with known true proportions, the RMSE of `SplitStrains` is less than 5%, as also shown in Table 1 below.

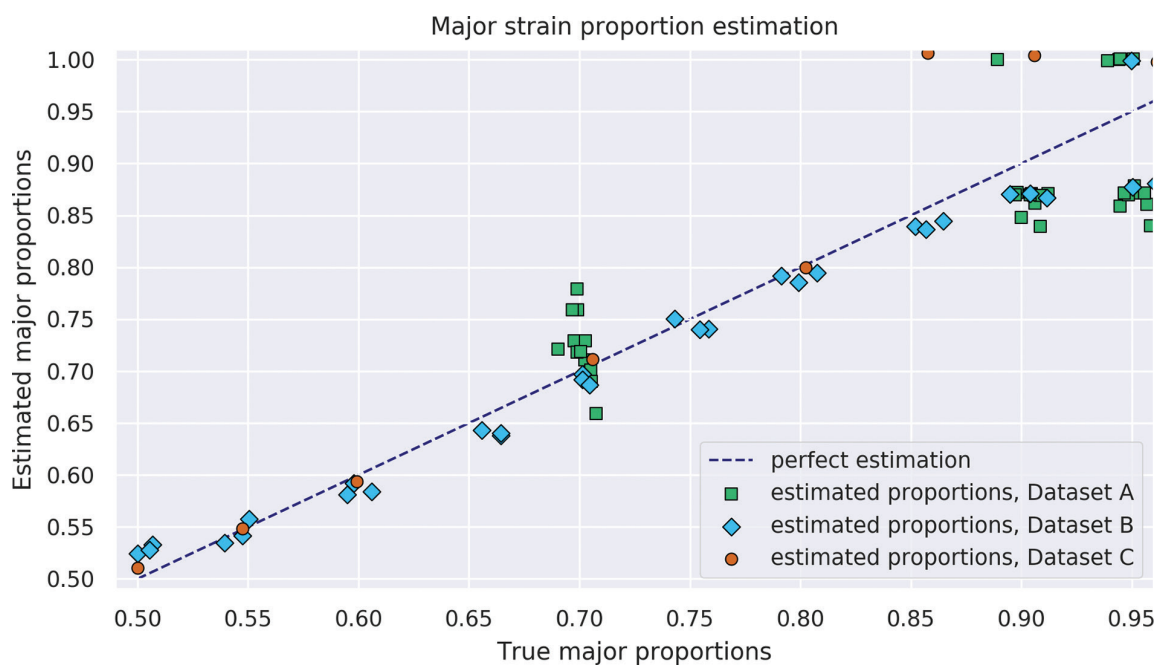

**Fig. 4.** Proportion estimation. The 74 true mixed samples and their major proportion estimated by `SplitStrains`.

**Table 1.** RMSE comparison across all datasets

| Dataset | Size (two-strains only) | SplitStrains | Mixed infection estimator | MixInfect | QuantTB |
|---------|------------------------|--------------|---------------------------|-----------|---------|
| A | 36 | **0.056** | 0.068 | 0.178 | 0.153 |
| B | 30 | 0.025 | **0.018** | 0.031 | 0.202 |
| C | 8 | 0.066 | 0.066 | **0.041** | 0.312 |
| Combined | 74 | **0.047** | 0.053 | 0.126 | 0.196 |

### Assignment of reads to constituent strains – Dataset C

Once the mixture model parameters have been estimated, the algorithm assigns each read containing one or more variable sites to a constituent strain using a Naïve Bayes approach. Note that those reads that do not contain any variant sites or have zero map quality remain unassigned (i.e. we perform a partial, rather than complete, strain reconstruction). In Figs 5 and 6 we respectively present the two-strain and three-strain confusion matrices to show the performance of this assignment on Dataset C, which is designed in such a way that the provenance of each read is known. We set $\alpha = 0.05$ to resolve the mixed strains. As explained above, the two-strain mixed sample with 95% major strain proportion is misclassified as pure, so Fig. 5 only contains seven contingency matrices. Here, a read is deemed to be classified correctly when it is assigned to its own strain, and incorrectly otherwise. These figures suggest that our assignment accuracy decreases as the two major strain proportions get close to each other, for samples with both two as well as three strains.

In practice, even if major and minor strain proportions are well apart, say 70:30, each individual variant in a read alignment file rarely has a clean 70:30 allele split. Instead, a variant's allele proportions take on values which are approximately normally distributed with respective means 70 and 30. `SplitStrains` successfully handles such variants. However, if a variant comprised of two alleles has an allele with a frequency below a user-specified threshold (the default being 10%), such a variant is deemed to be noisy and is not processed.

Using the read assignments to the strains, the algorithm outputs a new alignment file for each strain. In order to further evaluate the accuracy of the assignment, we create a consensus sequence from each alignment file. We expect the consensus sequences to match the respective genomes of the constituent strains. As the genome of each constituent strain has the same number $N$ of base substitutions relative to the reference genome, due to the way they are generated, the consensus sequences can have between 0 and $N$ mismatches with the true sequences. In the case of the two-strain mixtures, our algorithm successfully separates the strains with major strain proportion varying from 55% to 90%. However, as the major strain proportion gets closer to 50%, correct assignment of reads becomes steadily more challenging, as shown in Fig. 7. We note that SNPs in the reads that the alignment algorithm was unable to find a reasonable quality alignment for are not counted toward the errors in this analysis. Such reads are contained in the intervals [400130, 401700], [888990, 891520] and [2550020, 2551390] in the reference genome for *M. tuberculosis*, H37Rv (accession number NC 000962.3 at NCBI [12]), which fall within repetitive or mobile genome regions and are known to cause poor read alignments.

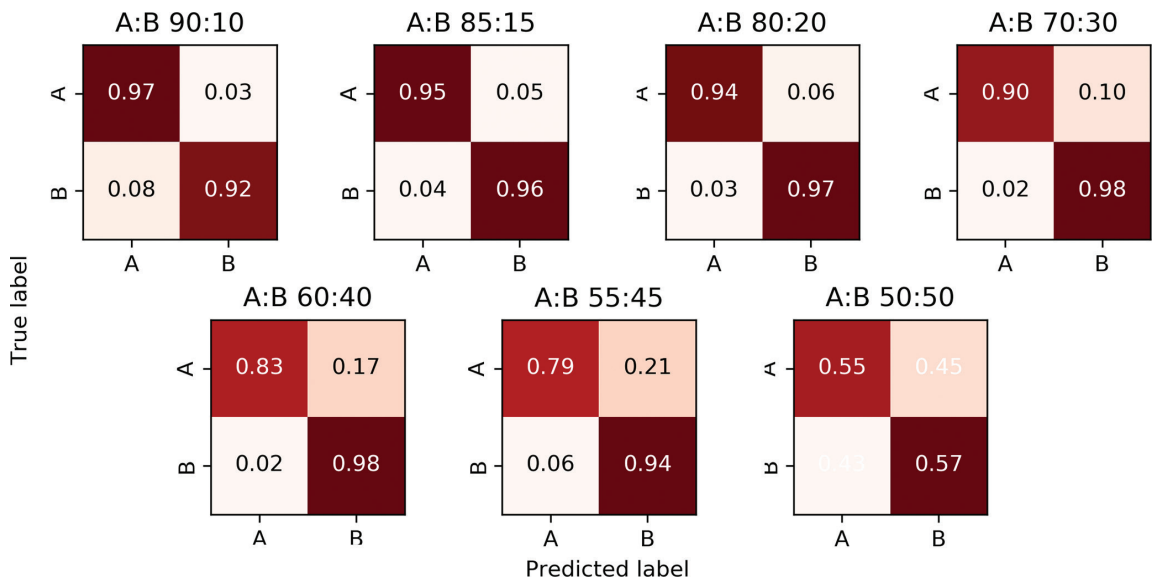

**Fig. 5.** Confusion matrices for two-strain samples, Dataset C. The major and minor strains are denoted A and B; their proportions are displayed above each matrix.

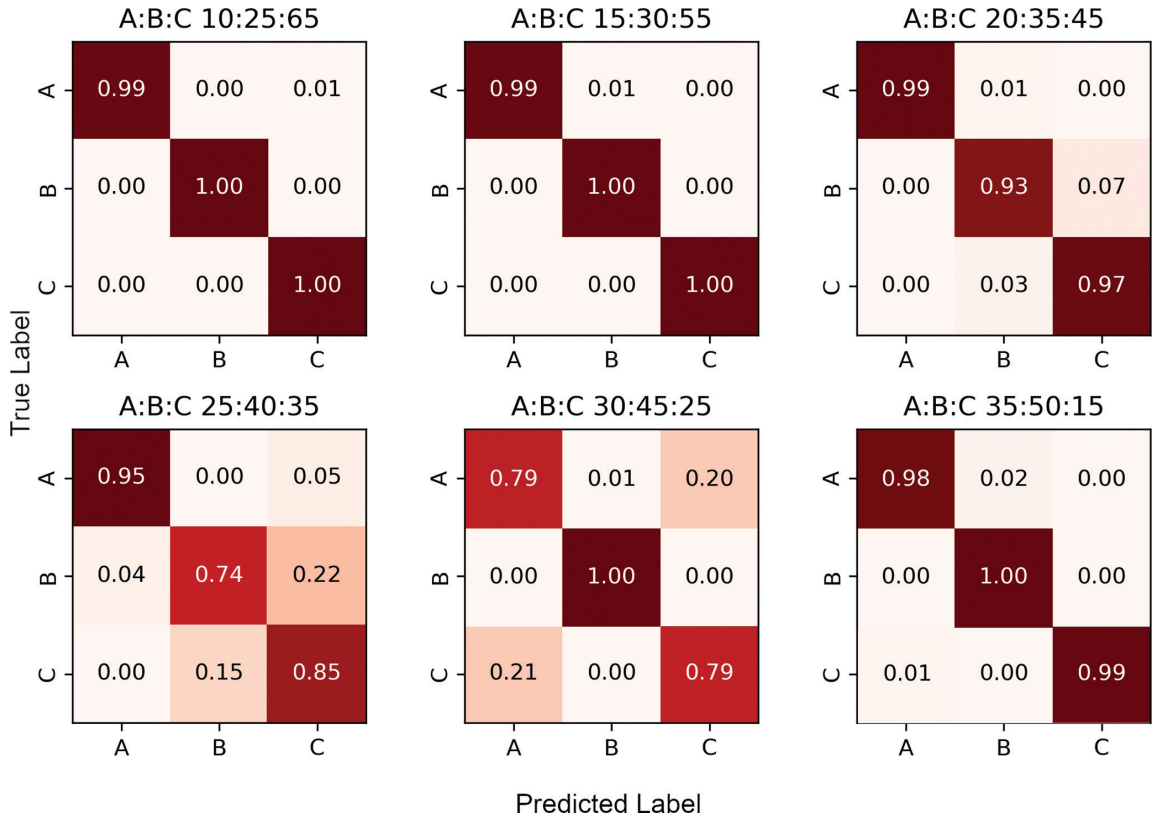

**Fig. 6.** Confusion matrices for three-strain samples, Dataset C. The strains in each sample are denoted A, B and C; their proportions are displayed above each matrix.

**Comparison with other tools**

`SplitStrains` consistently outperforms `MixInfect` [10], `QuantTB` [11] and `Mixed Infection Estimator` [2] in discriminating between pure and mixed infections. We illustrate their discrimination performance on datasets A, B and C combined in Fig. 8; their performance on each of the datasets separately is shown in Fig. 9 in the Supporting Information. The Receiver Operating Characteristic (ROC) curve shows the true positive rate (TPR) against the false positive rate (FPR) at various threshold settings. The Area Under the Curve (AUC) quantifies how well the algorithm is able to distinguish between pure and mixed infections. Higher AUC values mean that the algorithm is better at predicting the class of a sample. The statistics used as inputs to the AUC computation are as follows: likelihood ratio for `SplitStrains`, proportion het/total for `MixIn-fect`, deviance for `Mixed Infection Estimator`, and internal totscore for `Quant TB`. `SplitStrains` has the highest AUC (0.99) and can achieve close to 100% TPR with an FPR as low as 11%. The second best classification performance is obtained by `MixInfect`, with an AUC of 0.97. The ROC curves of `SplitStrains` and `MixInfect` are fairly close to one another, but the latter produces more false positives at higher true positive rates, giving a lower overall area under the curve.

`SplitStrains` also consistently obtains the lowest or second lowest proportion estimation error among the tools, and has the lowest error on the combined dataset. These results are shown in detail in Fig. 10 and summarized in Table 1.

**Performance with low genetic distance and depth of coverage**

We test the performance of `SplitStrains` on Dataset E which contains 32 pure and 32 mixed strains with small genetic distances (20, 30, 40 and 50 SNPs) and a low depth of coverage of 60. `SplitStrains`'s ability to detect mixed infections in this setting depends on the threshold $\alpha$. Higher $\alpha$ values allow the detection of mixed strains with small genetic distances at the risk of classifying pure strains as mixed.

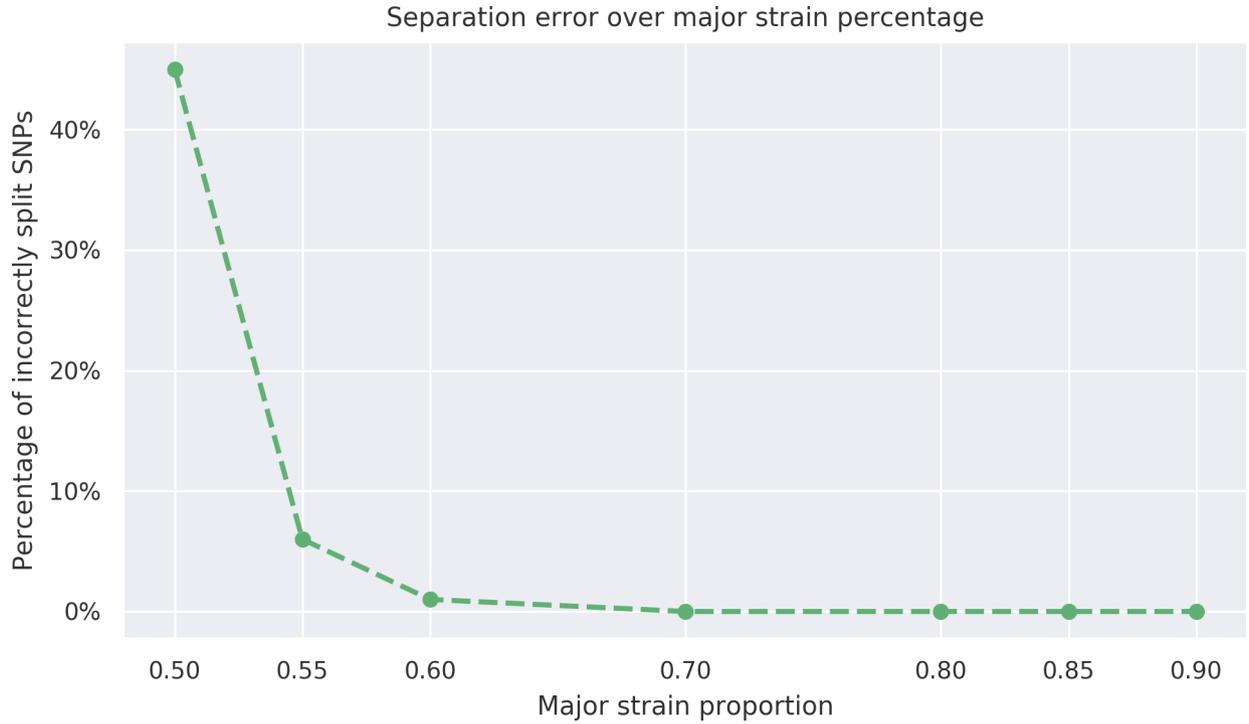

**Fig. 7.** Assignment error. The proportion of mismatches due to the incorrect assignment of reads among the positions where the strains differ from one another.

Fig. 11 shows the minimum $\alpha$ threshold needed to correctly identify mixed samples as mixed at a given SNP distance and major strain proportion while also correctly classifying

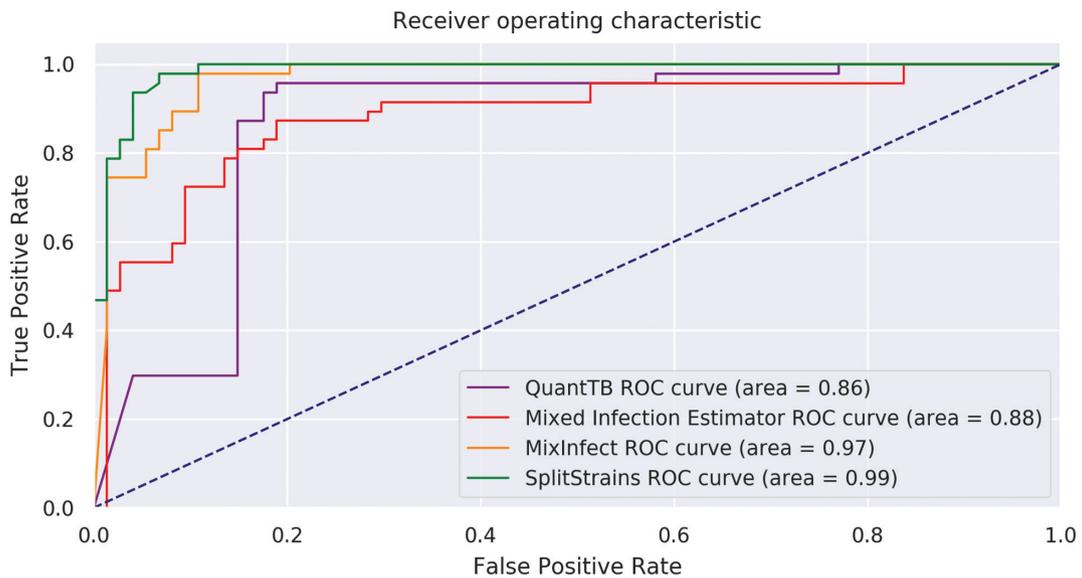

**Fig. 8.** ROC curves for the four tools. Performance in separating pure and mixed samples in datasets A, B and C.

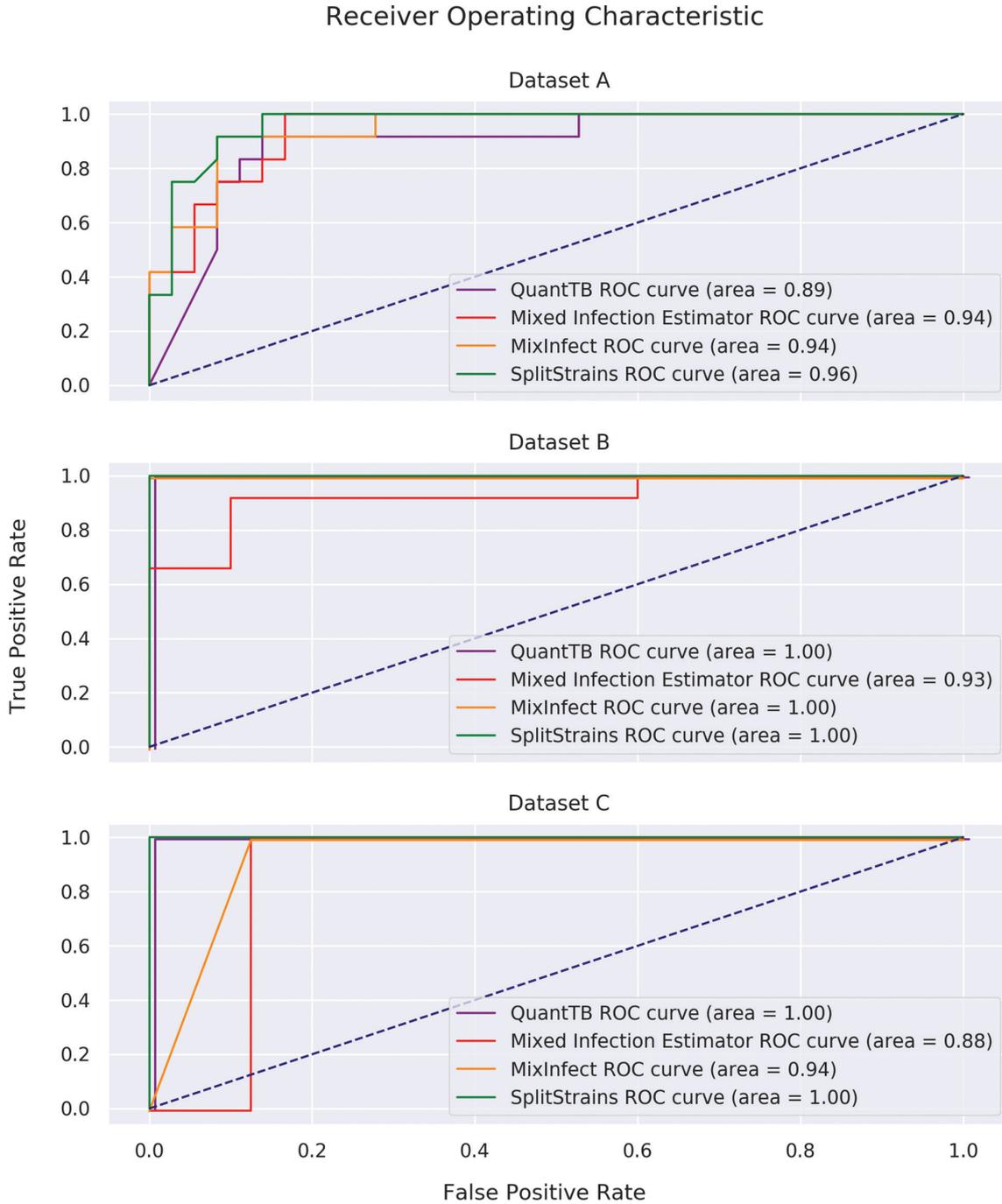

**Fig. 9.** ROC curves of all tools for each individual dataset A, B and C.

the pure samples as pure at the same SNP distance from the reference genome.

Fig. 11 shows that smaller genetic distances and larger major proportions require larger values of $\alpha$. For instance, a distance of 20 SNPs and a major strain proportion of 90% represents a challenging case which requires $\alpha$ to be set to 0.75, while the default value, $\alpha = 0.05$, suffices for a distance of 50 SNPs and a 50% major strain proportion.

We also show a separate ROC curve for Dataset E, in Fig. 12. Despite the challenges of classifying this Dataset, `Split-Strains` displays a reasonable trade-off between true positive and false positive rates and achieves an AUC of 0.96.

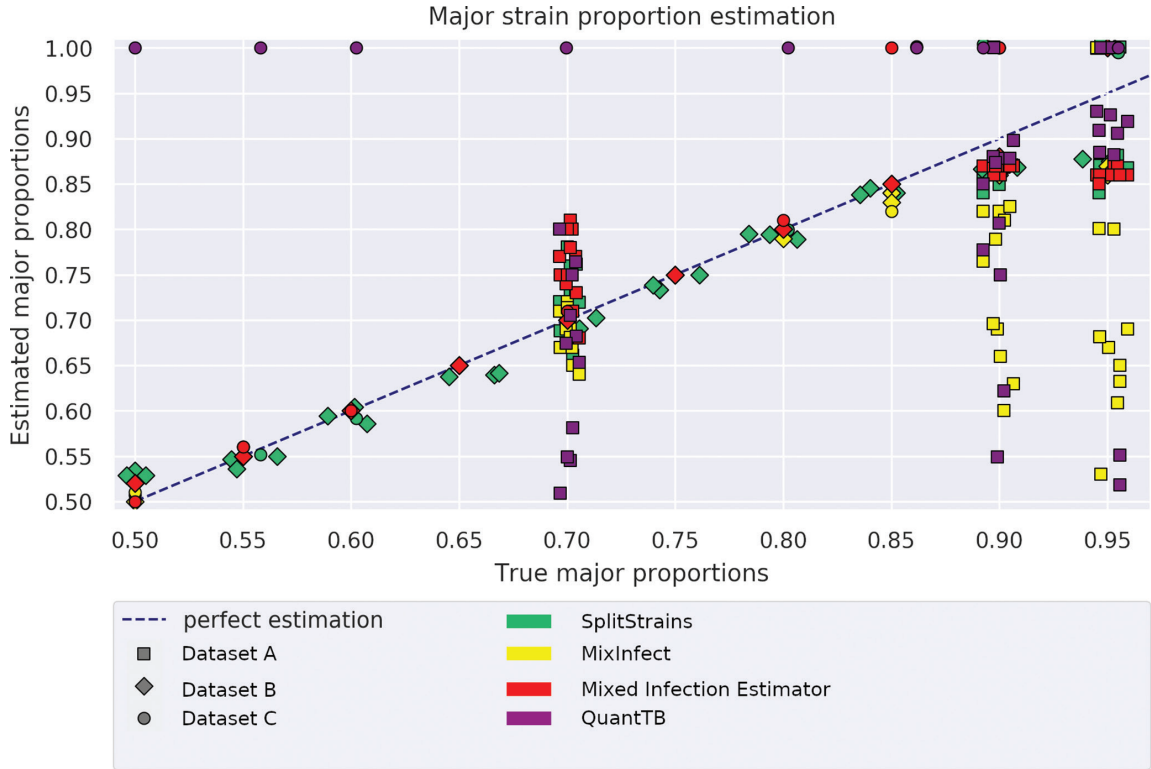

**Fig. 10.** Proportion estimation comparison. Major proportion estimates by four different tools on the 74 mixed samples from datasets A, B, and C.

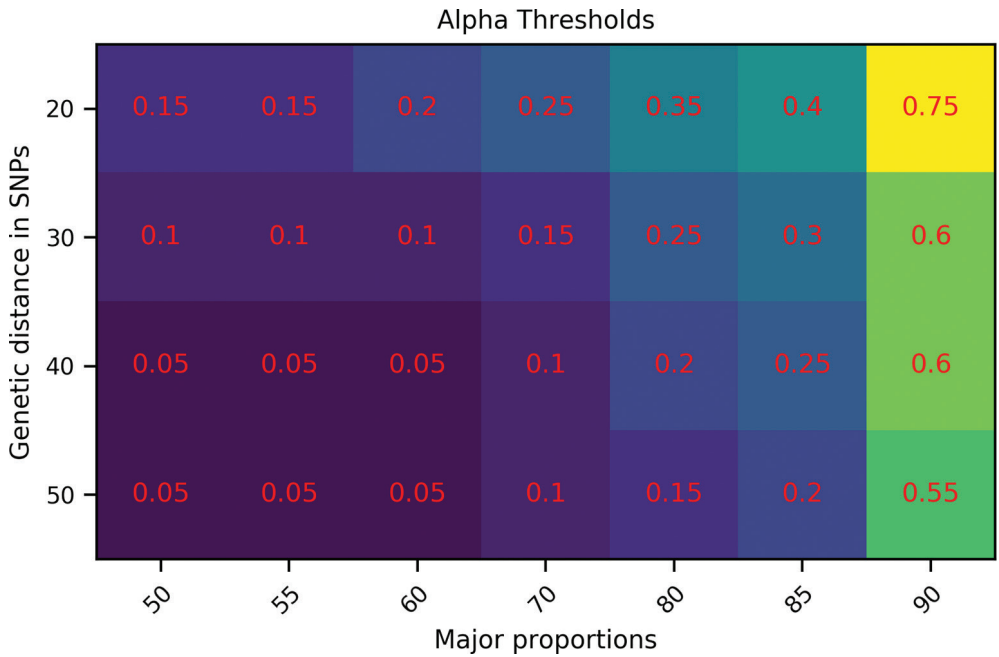

**Fig. 11.** α calibration, Dataset E. The α values required for the detection of mixed strains.

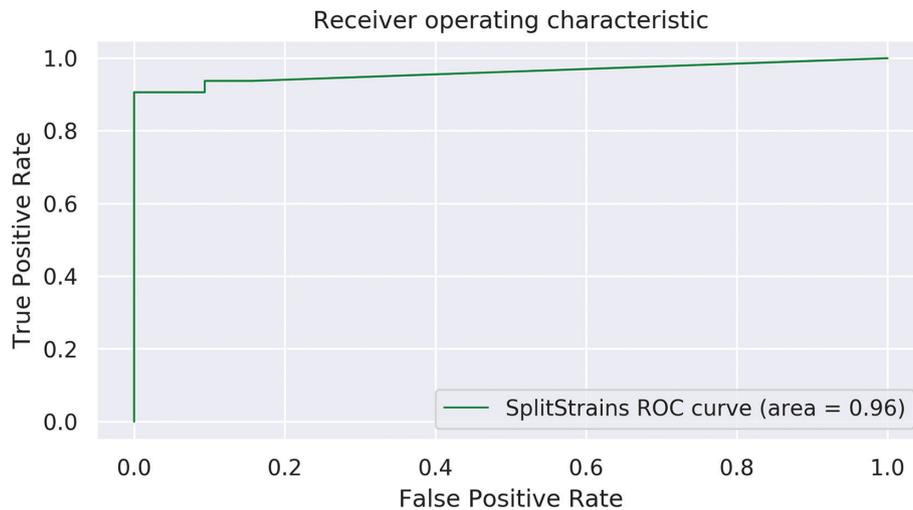

**Fig. 12.** ROC curve, Dataset E. The ROC curve for all 32 mixed and 32 pure samples.

## DISCUSSION

In this paper we introduced a novel algorithm, called `SplitStrains`, based on a rigorous statistical framework, for detecting multiple-strain infections, estimating the proportion of the major and minor strains, and partially reconstructing their sequences by assigning the reads that contain variants to one of these strains.

The *M.tuberculosis*'s genome is 4.4 million base pairs long and encodes roughly 4000 genes [27]. The genome features low amounts of recombination [28]. In addition, it has a high G+C content, with nearly two-thirds of all the base pairs being G-C [27]. This makes `SplitStrains` not only applicable to *M. tuberculosis*, but to other clonal pathogens whose genomic diversity is low and in the absence of significant amounts of recombination. This is the case of genetically monomorphic pathogens like *Yersinia pestis*, *Salmonella typhi*, and *Bacillus anthracis* [29], and pathogens that exhibit clonal evolution during infections [30]. It is also possible that `SplitStrains` could be applied to more recombinogenic bacteria like *Neisseria meningitidis* [31], provided that the recombination hotspots are removed in a preprocessing step using a software such as ClonalFrameML [32].

Although successful in many simulation scenarios, our algorithm suffers from the following limitations:

- When the proportions of two of the strains are very close to one another (for instance, a 50–50 mixture of two strains), the estimation becomes numerically sensitive and the identified strain sequences cannot be reliably inferred.
- When the overall depth of coverage is too low (below 20 or so), the proportions and the strain sequences cannot be reliably estimated either.
- The multiple infection status cannot be determined with confidence when the proportion of one of the strains is

below 10%, which is therefore the resolution of our algorithm.

- Although there is no required minimum threshold on genetic distance between the strains in a mixed sample, those with a high genetic similarity (around 20 SNPs apart) are challenging to identify and deconvolute. This suggests that our method will likely perform better on mixed samples due to multiple infection events than those due to within-host evolution [1].

We now discuss these limitations in turn, and sketch possible ways to improve on them. The situation when two strains are present at equal or near-equal proportions is rare in practice, since the strains will in general have unequal fitness and one will tend to dominate the other [5]. However, if it does occur, `SplitStrains` provides a useful diagnostic in the proportions it returns, and this can then serve as a starting point for separately culturing and sequencing each of the strains.

The situation when the depth of coverage is too low also occurs infrequently in practice, since most modern sequencing experiments tend to have a depth of coverage of at least 75 to 100. In the case of a low depth of coverage, for instance in a historical sequencing experiment, it is frequently not possible to reliably determine the sequence of even a simple strain, due to the difficulty in differentiating between a true SNP and a sequencing error. As can be expected, this difficulty is exacerbated in the presence of multiple strains. A more time-consuming method, such as a de Bruijn graph-based assembly, may be able to address this limitation in some situations [33].

Similarly, a very low-frequency minor strain is typically difficult to distinguish from noise due to sequencing errors, and in a situation when this appears to happen, a more targeted approach such as amplicon sequencing may be used to establish the sequence of at least a subset of the important genes. This approach may be used, for instance, to determine the resistance to a specific drug of the major and the minor strain [34].

Lastly, mixed samples with a small SNP distance (around 20) between the constituent strains should arise primarily through within-host evolution, although there is also a small likelihood of reinfection by two very similar strains in a high-prevalence region. Such similar strains may be more easily detectable by amplicon sequencing as well, provided that the sequencing is focused on the regions containing the variable positions. Alternatively, as we show in Fig. 11, the $\alpha$ significance level threshold in the `SplitStrains` algorithm can be increased to enable their detection in situations where such occurrences are expected to be frequent and the downside of false positive mixed calls is lower.

In spite of these limitations, we believe that our approach is a promising way forward, as also demonstrated by its favourable performance relative to existing methods. In particular, our approach appears to perform better than both `QuantTB` as well as `Mixed Infection Estimator`, and performs comparably to `MixInfect`, in most simulated settings, with regards to the task of identifying the presence of mixed infections and estimating proportions.

In addition, `SplitStrains` is unique among existing methods in its ability to provide additional information, namely, the assignment of each read to one of the underlying strains, with a subsequent identification of their sequence if desired. Importantly, like `QuantTB`, it does not rely on the knowledge of a large number of previously identified sequences, which is a clear advantage when investigating either a novel outbreak or an isolate originating from a data-poor setting. Furthermore, `SplitStrains` returns not only a call, but also a likelihood ratio, which is an indicator of the algorithm's confidence about the presence or absence of a mixed infection. We believe that, in situations where such information has either clinical or public health importance, the `SplitStrains` method will be a valuable addition to the existing collection of tools.

In future work, we plan to extend `SplitStrains` to work with other bacterial pathogens as well as to improve its resolution, at least in datasets with a high depth of coverage. Lastly, we plan to use `SplitStrains` as a pre-processing step in two pipelines - one for identifying related isolates in an outbreak [35], where mixed infections can mask such relatedness, and another one for predicting drug resistance [36], where mixed infections can impede a correct prediction when only the minor strain is drug-resistant.

**Funding information**
This work has been funded in part by a CANSSI Collaborative Research Team grant, 'Statistical methods for challenging problems in public health microbiology' and a Genome Canada grant, 'Machine Learning Methods to Predict Drug Resistance in Pathogenic Bacteria'. LC acknowledges funding from a Sloan Foundation fellowship (FG-2016-6392) and the MRC Centre for Global Infectious Disease Analysis (reference MR/R015600/1), jointly funded by the UK Medical Research Council (MRC) and the UK Foreign, Commonwealth and Development Office (FCDO), under the MRC/FCDO Concordat agreement, and is part of the EDCTP2 programme supported by the European Union. ML acknowledges funding from a NSERC Discovery grant.

**Acknowledgements**
The authors would like to acknowledge the invaluable input of Dr Cedric Chauve, Dr Theodore Cohen, Dr John Lees, Dr Nicholas Croucher, and Hooman Zabeti. They would also like to thank Dr Patrick Cudahy for beta-testing an earlier version of the tool.

**Author contributions**
E. G., methodology, software, validation, formal analysis, visualization, writing – original draft preparation, writing – review and editing. M. M. M., data curation, resources, writing – review and editing. I. C., data curation, resources, supervision, writing – review and editing. M. L., conceptualisation, methodology, formal analysis, supervision, funding, writing – review and editing. L. C., conceptualisation, methodology, formal analysis, supervision, project administration, writing – original draft preparation, writing – review and editing.

**Conflicts of interest**
The authors declare that there are no conflicts of interest.

**References**
1. Cohen T, Helden PD, Wilson D, Colijn C, McLaughlin MM, *et al.* Mixed-strain *Mycobacterium tuberculosis* infections and the implications for tuberculosis treatment and control. *Clin Microbiol Rev* 2012;25:708–719.
2. Eyre DW, Cule ML, Griffiths D, Crook DW, Peto TEA, *et al.* Detection of mixed infection from bacterial whole genome sequence data allows assessment of its role in *Clostridium difficile* transmission. *PLoS Comput Biol* 2013;9:e1003059.
3. Shaj, Almagro G, Mc V. Deconvolution of multiple infections in *Plasmodium falciparum* from high throughput sequencing data. *Bioinformatics* 2017;34:9–15.
4. Nathavitharana RR, Shi CX, Chindelevitch L, Calderon R, Zhang Z, *et al.* Polyclonal pulmonary Tuberculosis infections and risk for multi-drug resistance, LIMA, Peru. *Emerg Infect Dis* 2017;23:1887–1890.
5. Sergeev R, Colijn C, Cohen T. Models to understand the population-level impact of mixed strain *M. Tuberculosis* infections. *J Theor Biol* 2011;280:88–100.
6. Weiss S, David S, Victor I. Heteroresistance: A cause of unexplained antibiotic treatment failure? *PLOS Pathogens* 2019;15:1–7.
7. Zong Z, Huo F, Shi J, Jing W, Ma Y, *et al.* Relapse versus reinfection of recurrent tuberculosis patients in a national Tuberculosis specialized hospital in Beijing, China. *Front microbiol* 2018;9:1858.
8. Nadon CA, Trees E, Ng L, Møller E, Reimer A, *et al.* Development and application of MLVA methods as a tool for inter-laboratory surveillance. *Euro Surveill* 2013;18.
9. Leonid C, Colijn C, Moodley P, Wilson D, Cohen T, *et al.* ClassTr: Classifying within-host heterogeneity based on tandem repeats with application to *Mycobacterium tuberculosis* infections. *PLOS Computational Biology* 2016;12:1–16.
10. Sobkowiak B, Glynn JR, Houben RMGJ, Mallard K, Phelan JE, *et al.* Identifying mixed *Mycobacterium tuberculosis* infections from whole genome sequence data. *BMC Genomics* 2018;19:613.
11. Anyansi C, Keo A, Walker BJ, Straub TJ, Manson AL, *et al.* QuantTB - a method to classify mixed *Mycobacterium tuberculosis* infections within whole genome sequencing data. *BMC genomics* 2020;21:80.
12. O'Leary NA, WrightM, Brister R, Ciufo S, Haddad D, *et al.* Reference sequence (Refseq) database at NCBI: Current status, taxonomic expansion, and functional annotation. *Nucleic Acids Res* 2016;44:D733–D745.
13. Moon TK. The expectation-maximization algorithm. *IEEE Signal Process Mag* 1996;13:47–60.
14. Feijao P, Yao H-T, Fornika D, Gardy J, Hsiao W, *et al.* MentaLiST – A fast MLST caller for large MLST schemes. *Microb Genom* 2018;4.
15. De J, Michael A, Gerrick E, Xu W, Park S, *et al.* Comprehensive essentiality analysis of the *Mycobacterium tuberculosis* genome via saturating transposon mutagenesis. *MBio* 2017;8:16–e02133.

16. **Huang W**, **Li L**, **Myers JR**, **Marth GT**. Art: A next-generation sequencing read simulator. In: *Bioinformatics*, Vol. 28. Oxford University Press, 15 2012. 2012. pp. 593–594.

17. **Li H**. Aligning sequence reads, clone sequences and assembly contigs with BWA-MEM. *arXiv* 2013:1303.3997.

18. **Bolger A**, **Lohse M**, **Usadel B**. Trimmomatic: A flexible trimmer for Illumina sequence data. *Bioinformatics* 2014;30:2114–2120.

19. **Li H**, **Handsaker B**, **Wysoker A**, **Fennell T**, **Ruan J**, *et al*. The sequence Alignment/map format and samtools. *Bioinformatics* 2009;25:2078–2079.

20. **Comas I**, **Jaidip C**, **Peter MS**, **James G**, **Stefan N**, *et al*. Human T cell epitopes of *Mycobacterium tuberculosis* are evolutionarily hyper-conserved. *Nat Genet* 2010;42:498–503.

21. **Virtanen P**, **Travis O**, **Matt H**, **Tyler R**, **David C**, *et al*. SCIPY 1.0: Fundamental algorithms for scientific computing in Python. *Nat Methods* 2020;17:261–272.

22. **Zignol M**, **Andrea C**, **Anna SD**, **Philippe G**, **Natavan A**, *et al*. Genetic sequencing for surveillance of drug resistance in tuberculosis in highly endemic countries: A multi-country population-based surveillance study. *Lancet Infect Dis* 2018;18:675–683.

23. **J A**, **Crampin AC**, **Houben RM**, **Mzembe T**, **Mallard K**, *et al*. Large-scale whole genome sequencing of M. Tuberculosis provides insights into transmission in a high prevalence area. Guerra-assunção. *elife* 2015;4:e05166.

24. **Forouzan E**, **Parvin S**, **Masoumeh SM**, **Karkhane AA**, **Yakhchali B**, *et al*. Practical evaluation of 11 de novo assemblers in metage-nome assembly. *J Microbiol Met* 2018;151:99–105.

25. **Wajid B**, **Serpedin E**. Review of general algorithmic features for genome assemblers for Next Generation sequencers. *Genomics, Proteomics & Bioinformatics* 2012;10:58–73.

26. **Goig GA**, **Silvia B**, **Alberto L**, **Garcia B**, **Iñaki C**, *et al*. Contaminant dna in bacterial sequencing experiments is a major source of false genetic variability. *BMC Biol* 2020;18:1–15.

27. **Cole S**, **Churcher C**, **Parkhill J**, **Garnier T**, **Harris D**, *et al*. Deciphering the biology of *Mycobacterium tuberculosis* from the complete genome sequence. *Nature* 1998;396:190.

28. **Chiner-Oms Á**, **Sánchez-Busó L**, **Corander J**, **Gagneux S**, **Harris SR**, *et al*. Genomic determinants of speciation and spread of the *Mycobacterium tuberculosis* complex. *Sci Adv* 2019;5:eaaw3307.

29. **Achtman M**. Insights from genomic comparisons of genetically monomorphic bacterial pathogens. In: *The Royal Society, 2012, Philosophical Transactions of the Royal Society B: Biological Sciences*, Vol. 367. 2012. pp. 860–867.

30. **Didelot X**, **Walker AS**, **Peto TE**, **Crook DW**, **Wilson DJ**, *et al*. Within-host evolution of bacterial pathogens. In: *Nature Reviews Microbiology*, Vol. 14. Nature Publishing Group, 2016. pp. 150–162.

31. **Vos M**, **Didelot X**. A comparison of homologous recombination rates in bacteria and archaea. *ISME J* 2009;3:199–208.

32. **Didelot X**, **Wilson DJ**. Clonalframeml: Efficient inference of recombination in whole bacterial genomes. *PLoS Comput Biol* 2015;11:e1004041.

33. **Holley G**, **Melsted P**. Bifrost – highly parallel construction and indexing of colored and compacted de Bruijn graphs. *bioRxiv* 2019.

34. **Colman RE**, **Schupp JM**, **Hicks ND**, **Smith DE**, **Buchhagen JL**, *et al*. Detection of low-level mixed-population drug resistance in *Mycobacterium tuberculosis* using high fidelity amplicon sequencing. *PLoS One* 2015;10:e0126626.

35. **Katebi M**. In: *Pathogist: a Novel Method for Clustering Pathogen Isolates by Combining Multiple Genotyping Signals*. Simon Fraser University, 2019.

36. **Zabeti H**. *An Interpretable Classification Method for Predicting Drug Resistance in M. tuberculosis*. Cold Spring Harbor Laboratory, bioRxiv, 2020.